# Performance Enhancement of Black Phosphorus Field-Effect Transistors by Chemical Doping

Yuchen Du, Lingming Yang, Hong Zhou, and Peide D. Ye, *Fellow, IEEE*

*Abstract*— In this letter, a new approach to chemically dope black phosphorus (BP) is presented, which significantly enhances device performance of BP field-effect transistors (FETs) for an initial period of 18 h, before degrading to previously reported levels. By applying 2,3,5,6-tetrafluoro-7,7,8,8-tetracyanoquinodimethane (F4-TCNQ), low ON-state resistance of 3.2 Ω·mm and high field-effect mobility of 229 cm$^2$/Vs are achieved with a record high drain current of 532 mA/mm at a moderate channel length of 1.5 *μm*.

*Index Terms*— Black phosphorus, contact resistance, MOSFET, phosphorene, semiconductor device doping.

## I. INTRODUCTION

BLACK phosphorus (BP) is a stable phosphorus allotrope at room temperature. The bandgap in BP is dependent on the number of layers, ranging from 0.3 eV in bulk material to 2.0 eV in monolayer phosphorene [1]. From the transport perspective, the hole mobility in BP is on the order of 10$^4$ cm$^2$/Vs at cryogenic temperatures [2], having the potential to outperform traditional silicon or III-V semiconductors at the ballistic limit [3]. The presence of a sizable bandgap along with high carrier mobility makes BP a competitive material for future electrical applications [4]-[7]. To date, experimental demonstrations of few-layer BP FETs have been achieved by several groups [8]-[10], and sustain good electrical performance for weeks or months with different passivations techniques [11]-[13], functioning even at radio frequencies [14], [15]. In addition, BP devices are also suitable for optoelectronic applications [16]-[20]. More encouragingly, recent work based on few-layer phosphorene encapsulated by atomically thin hexagonal boron nitride has allowed the mobility to approach 6000 cm$^2$/Vs at low temperatures [21], and Shubnikov-de Haas oscillations and quantum Hall effects have been observed [21]-[26]. Since all these results were based on few-layer phosphorene as part of the metal-oxide-semiconductor (MOS) structure, improving the quality of such structure can have broad impacts on future developments. In this letter, for the first time, we explored F4-TCNQ doping to enhance BP FET device performance. The simplicity and efficiency of this doping technique present a valuable approach to achieve high-performed BP thin-film transistors.

## II. EXPERIMENT

Few-layer BP flakes were exfoliated from bulk crystal, and

Yuchen Du, Lingming Yang, Hong Zhou, and Peide D. Ye are with the School of Electrical and Computer Engineering and Birck Nanotechnology Center, Purdue University, West Lafayette, IN 47907 USA (e-mail: yep@purdue.edu). The work is supported by NSF under Grant ECCS-1449270, NSF/AFOSR EFRI 2-DARE Program, and ARO under Gant W911NF-14-1-0572.

then transferred onto a heavily doped silicon substrate covered with a 90 nm SiO$_2$ layer. Subsequently source/drain metal contacts were formed by photolithography and a 70 nm thick Ni metal deposition and lift-off process. Electrical measurements were first carried out on the completed MOSFET devices under an ambient atmosphere using a Keithley 4200 semiconductor parameter analyzer. The measured sample was then soaked in an isopropyl alcohol solution of 0.75 mmol/L F4-TCNQ for 30 min, followed by a spin-drying cycle at 2000 rpm for 10 s. No further N$_2$ blow or heat-drying was applied after spin-drying. F4-TCNQ is an organic molecule with a large electron affinity of 5.2 eV. It is widely used as a p-type dopant for carbon nanotube and graphene applications [27]-[29]. After doping, the BP FET was electrically characterized again under ambient condition. A schematic view and optical micrograph of a doped BP FET with 1.5 *μm* channel length are illustrated in Fig. 1(a) and 1(b), respectively. The thickness of BP flake for this particular device was measured by atomic force microscopy (AFM) to be 8.4 nm as shown in Fig. 1(c).

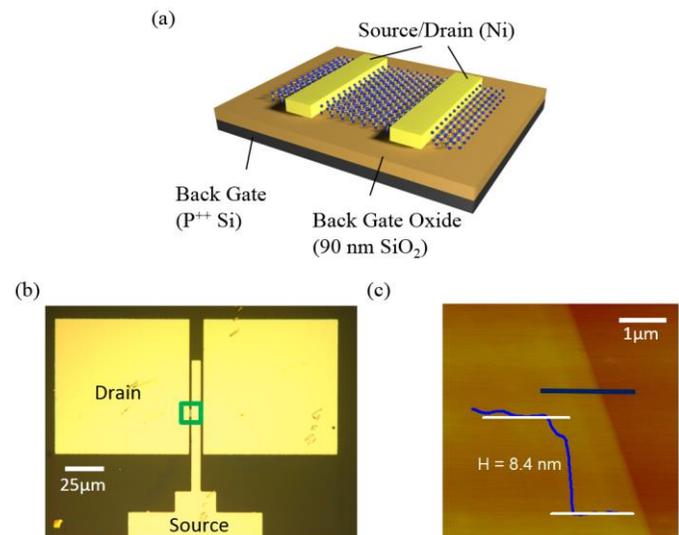

Fig. 1 (a) Schematic view of device structure. (b) Optical image of doped BP FET with a channel length of 1.5 *μm* defined by photolithography. The green box highlights the device location, with source and drain regions labeled. (c) AFM trace shows the BP film is 8.4 nm thick.

## III. RESULT AND DISCUSSION

A clear signature of p-type transistor behavior can be seen on the transfer characteristics of the few-layer BP FET, as shown in Fig. 2(a). Inspecting the transfer characteristic of the p-type BP transistor, we can estimate the field-effect mobility according to the square law theory. The extrinsic field-effect mobility before F4-TCNQ doping is 181 cm$^2$/Vs. After doping, the on-state current increased significantly. More importantly, the threshold voltage ($V_T$), extracted from linear extrapolation at low $V_{ds}$ bias, shifts in the direction of positive gate voltage.

The $V_T$ shift can be explained by the move of the Fermi level towards the valence band edge of the semiconducting BP due to the p-type chemical doping, where the F4-TCNQ on the surface of BP absorbs electrons and generates holes in the channel [27]-[29]. The extrinsic field-effect mobility after F4-TCNQ is estimated to be 229 cm$^2$/Vs. Such a 30% increase in field-effect mobility after F4-TCNQ is associated with the reduction of surface scattering and the immunity of flake quality degradation in the channel. F4-TCNQ, an organic doping layer spin-coated on the surface of BP, not only is an effective passivation layer to avoid flake degradation to ambient exposure, but also prevents surface scattering from neighboring adsorbates. However, a decreased on/off ratio from $10^3$ to 10 was also observed. Part of the reason for this low on/off ratio could be the leakage current through the F4-TCNQ film or along the F4-TCNQ/SiO$_2$ interface. This suggests that the local doping in a specific device area should be studied in future work. The drain current varies linearly with small source/drain biases as shown in characteristics of Fig. 2(b) and 2(c), implying an ohmic-like contact resistance at the Ni/BP interface with a small Schottky barrier. Before doping, the maximum drain current at $V_{ds}$ = -2 V is 211 mA/mm, which is already good performance for a device with a micrometer channel length. After F4-TCNQ doping, the maximum drain current is more than doubled to 532 mA/mm, and the on-state resistance is reduced from 7.4 Ω·mm to 3.2 Ω·mm. To the best of our knowledge, this doped BP FET with bulk global gate has demonstrated the lowest on-state resistance and highest drain current compared to those with similar or even shorter channel length [8-11, 14, 15].

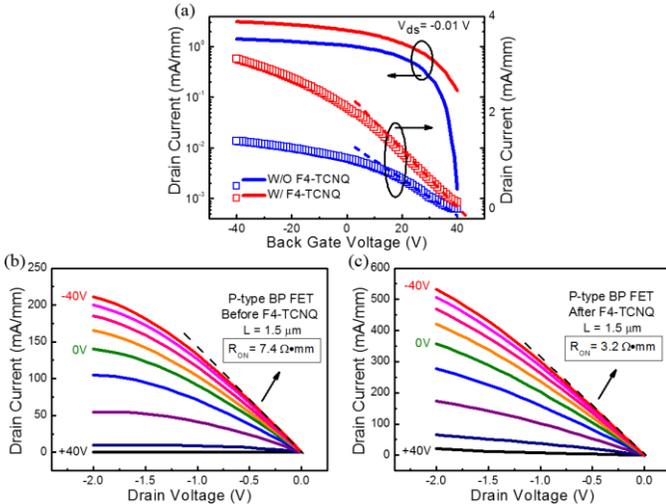

Fig. 2 (a) Transfer characteristics of a p-type few-layer BP FET before and after F4-TCNQ doping. Solid curves refer to the semi-logarithmic scale on the left, and square symbols refer to the linear scale on the right. All data are taken at a constant drain bias of -0.01 V. Threshold voltages for both without and with F4-TCNQ doping can be determined by the dashed lines, to be 38 V and 43 V, respectively. Output characteristics of the p-type BP FET before (b) and after (c) F4-TCNQ.

In order to investigate the effects of F4-TCNQ chemical doping in greater detail, we have measured sheet resistance and contact resistance separately utilizing the transfer length method (TLM) according to,

$$R_{total} = 2R_c + R_s \frac{L}{W} \quad (1)$$

where $R_{total}$ is the total resistance, $R_c$ is contact resistance, $R_s$ is sheet resistance, $L$ is channel length, and $W$ is channel width. An example of $R_{total}$ versus $L/W$ is shown in Fig. 3(a), where TLM resistances of BP FETs before and after F4-TCNQ doping are extracted at a constant back gate voltage of -40 V. Significant reduction in the slope of $R_{total}$ can been seen after doping, indicating a remarkable reduction in sheet resistance. Sheet resistances at different back gate voltages are plotted in Fig. 3(b). A high bias region with small absolute error bars has been chosen to have an accurate comparison between sheet and contact resistances under the effects of F4-TCNQ doping. The sheet resistance decreases from 3.3 kΩ to 1.2 kΩ at $V_{bg}$= -40 V. Such 3 times reduction in sheet resistance further verifies the increase of hole concentration from doping since mobility does not increase dramatically. 2D carrier concentration has been determined from the equation,

$$n = \frac{L}{eWR_s\mu_{FE}} \quad (2)$$

where $e$ is electron charge, $R_s$ is sheet resistance, and $\mu_{FE}$ is intrinsic field-effect mobility. The 2D sheet density before and after doping are 7.5 × 10$^{12}$ cm$^{-2}$ and 2.2 × 10$^{13}$ cm$^{-2}$ at $V_{bg}$ = 0V, respectively, indicating a strong doping effect as a result of the F4-TCNQ layer. These values are consistent with the results from Hall measurements of similar films. Contact resistance has also been calculated and analyzed. A plot of contact resistance under varying bias is shown in Fig. 3(c), where normalized contact resistance measured before and after doping are 1.7 Ω·mm and 1.3 Ω·mm, respectively, at $V_{bg}$= -40V. This nearly 1.3 times reduction of contact resistance is ascribed to the shrinking of the Schottky barrier width by doping of the BP film underneath the contacts.

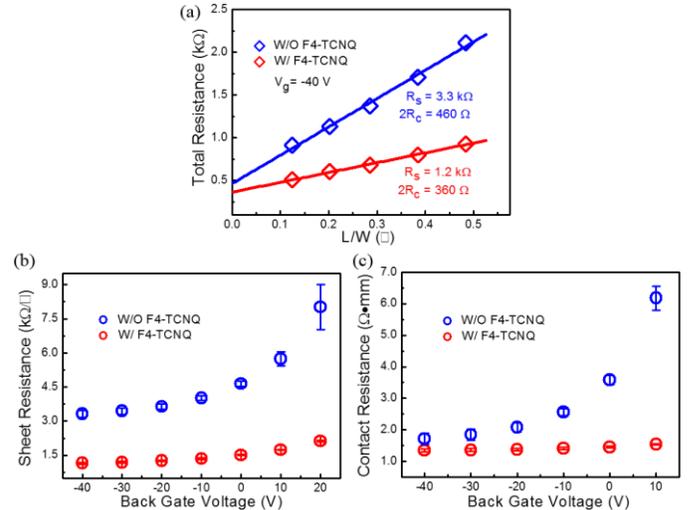

Fig. 3 (a) TLM resistance of BP FETs before and after F4-TCNQ doping at a constant back gate voltage of -40 V. The sheet resistances are extracted from the slopes of linear fitting, and the contact resistances are extracted from the intercepts of fitting curves. (b) Sheet resistance and (c) Contact resistance before and after doping vary with different back gate biases. Error bars are determined from the standard errors of the linear fitting.

A statistical study of key parameters is necessary and important so as to gain a comprehensive understanding of F4-TCNQ

doping effect on BP transistors. A total of five BP flakes with similar thickness of approximately 8.4 nm have been patterned with TLM structures. Over 30 devices have been carefully fabricated, systematically measured, and comprehensively analyzed. Fig. 4(a) shows the results of this study for sheet resistance. The sheet resistances before doping vary from 8.0 kΩ/□ to 3.2 kΩ/□, which is directly associated with anisotropic mobility in BP [1], [5], [6]. After doping, the sheet resistance is dramatically reduced in each case, with an average reduction of 2.5 times. Similarly, the contact resistance before and after F4-TCNQ doping for each TLM is reduced on average by 40% as shown in Fig. 4(b). We describe the doping efficiency in terms of on-resistance improvement factor (on-resistance before doping / on-resistance after doping) and mobility improvement factor (mobility after doping / mobility before doping) for each single device. A box chart of on-resistance factor versus different channel lengths is shown in Fig. 4(c). It should be noted that the doping efficiency decreases as the channel length scales from 3 $\mu m$ to 100 nm. This channel length dependent doping efficiency is ascribed to on-resistance saturation in the short channel regime [8]. Undeniably, doping has a much stronger effect on sheet resistance than contact resistance. As the channel length aggressively scales down, the contact resistance becomes the dominant factor, compared to sheet resistance. Therefore, the doping efficiency drops for shorter devices. Our experiments also show the on-state resistance factor has dropped from an average 2.2 times at a channel length of 3 $\mu m$ to 1.4 times at a channel length of 100 nm. In addition, the mobility factor's fluctuation with different channel lengths is shown in Fig. 4(d). The consistent doping efficiency with an average of 1.7 times enhancement in field-effect mobility presents a channel length independent behavior, and has further confirmed the effectiveness of F4-TCNQ doping on BP transistors.

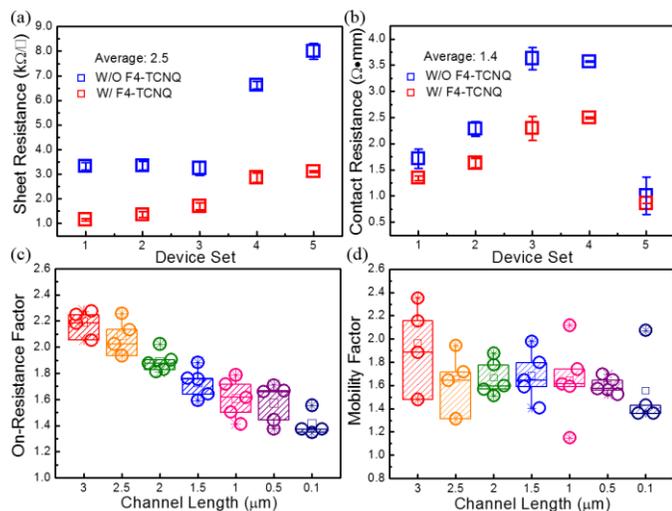

Fig. 4 Statistical studies of (a) Sheet resistance and (b) Contact resistance for different TLM. Error bars are determined from the standard errors of the linear fitting in each individual TLM structure. Doping effects on (c) On-state resistance and (d) Field-effect mobility for different channel lengths. (Circles stand for actual data, crosshatched boxes represent the interquartile range, stars indicate the maximum/minimum data point, and small squares point to the mean value).

The stability of F4-TCNQ doped BP transistors was also studied in our experiment, as shown in Fig. 5. The doped FETs have been stored, and measured repeatedly under ambient condition. Air-stability of a F4-TCNQ doped BP device was monitored by measuring the electrical performance as a function of time. Before doping, on-state resistance of the 1.5 $\mu m$ channel length device was 21.2 Ω·mm. The device was immediately measured again after the doping, and showed a significant drop in on-resistance, indicative of a successful doping. Meanwhile, the on-state current and off-state current also increased due to the charge transfer effect. By 18 hours after application of doping layer, the device has restored back to the original performance level, indicating a reversible capability of F4-TCNQ doping. Furthermore, a long-term reservation of device performance using F4-TCNQ doping on BP FETs needs to be explored in near future, for example by using a special polymer layer to cover the F4-TCNQ doping layer and prevent it from reacting with oxygen in its surrounding environment [30].

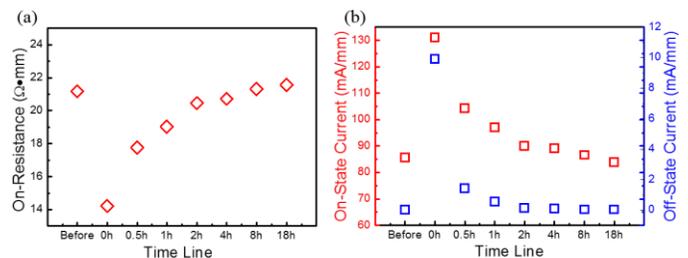

Fig. 5 (a) On-resistance increases with time. (b) On-state current (extracted at $V_{bg}$= -40 V, and $V_{ds}$= -2 V), shown in red square, and off-state current (extracted at $V_{bg}$= 40 V, and $V_{ds}$= -2 V) which is identified as blue square elapse with time.

## IV. CONCLUSION

In summary, we have characterized the chemical doping effect of a F4-TCNQ layer on BP FETs. The doping technique has enhanced the electrical performance of BP transistors with a record high drain current of 532 mA/mm. This p-type doping also presents an average of 2.5 times reduction in sheet resistance, 1.4 times decrease in contact resistance, and 1.7 times enhancement in field-effect mobility in statistics. Better doping techniques are extremely important for further reducing the contact resistance and enhancing the transport properties of materials, including BP. To develop different doping techniques in 2D materials with long-time stability remains a grand challenge.

## REFERENCES

[1] H. Liu, Y. Du, Y. Deng, and P. D. Ye, "Semiconducting Black Phosphorus: Synthesis, Transport Properties and Electronic Applications," *Chem. Soc. Rev.*, vol. 44, no. 9, pp. 2732-2743, May 2015. DOI: 10.1039/C4CS00257A
[2] A. Morita, "Semiconducting Black Phosphorus," *Appl. Phys. A: Mater. Sci. Process.*, vol. 39, no. 4, pp. 227–242, Apr. 1986. DOI: 10.1007/BF00617267
[3] K. Lam, Z. Dong, and J. Guo, "Performance Limits Projection of Black Phosphorus Field-effect Transistors," *IEEE Electron Device Lett.*, vol. 35, no. 9, pp. 963–965, Sept. 2014. DOI: 10.1109/LED.2014.2333368
[4] L. Li, Y. Yu, G. Ye, Q. Ge, X. Ou, H. Wu, D. Feng, X. Chen, and Y. Zhang, "Black Phosphorus Field-effect Transistors," *Nat. Nanotechnol.*, vol. 9, no. 5, pp. 372-377, Mar. 2014. DOI: 10.1038/nnano.2014.35
[5] H. Liu, A. T. Neal, Z. Zhu, Z. Luo, X. Xu, D. Tomanek, and P. D. Ye, "Phosphorene: An Unexplored 2D Semiconductor with a High Hole Mobility," *ACS Nano,* vol. 8, no. 4, pp. 4033-4041, Mar. 2014. DOI: 10.1021/nn501226z

[6] F. Xia, H. Wang, and Y. Jia, "Rediscovering Black Phosphorus as an Anisotropic Layered Material for Optoelectronics and Electronics," *Nat. Commun.*, vol. 5, no. 4458, pp. 4458-1-4458-6, Jul. 2014. DOI: 10.1038/ncomms5458

[7] A. Gomez, L. Vicarelli, E. Prada, J. Island, K. Acharya, S. Blanter, D.. Groenendijk, M. Buscema, G. Steele, J. Alvarez, H. Zandbergen, J. Palacios, and H. Zant, "Isolation and Characterization of Few-layer Black Phosphorus," *2D Mat.*, vol. 1, no. 2, pp. 025001, Jun. 2014. DOI: 10.1088/2053-1583/1/2/025001

[8] Y. Du, H. Liu, Y. Deng, and P. D. Ye, "Device Perspective for Black Phosphorus Field-Effect Transistors: Contact Resistance, Ambipolar Behavior, and Scaling," *ACS Nano*, vol. 8, no. 10, pp. 10035-10042, Oct. 2014. DOI: 10.1021/nn502553m

[9] N. Haratipour, M. Robbins, and S. Koester, "Black Phosphorus p-MOSFETs with 7-nm $HfO_2$ Gate Dielectric and Low Contact Resistance," *IEEE Electron Device Lett.,* vol. 36, no. 4, pp. 411–413, Apr. 2014. DOI: 10.1109/LED.2015.2407195

[10] S. Das, M. Demarteau, and A. Roelofs, "Ambipolar Phosphorene Field Effect Transistor," *ACS Nano*, vol. 8, no. 11, pp. 11730-11738, Oct. 2014. DOI: 10.1021/nn505868h

[11] W. Zhu, M. N. Yogeesh, S. Yang, S. H. Aldave, J.-S. Kim, S. Sonde, L. Tao, N. Lu, and D. Akinwande, "Flexible Black Phosphorus Ambipolar Transistors, Circuits, and AM Demodulator," *Nano Lett.*, vol. 15, no. 3, pp. 1883-1890, Feb. 2015. DOI: 10.1021/nl5047329

[12] J. Wood, S. Wells, D. Jariwala, K. Chen, E. Cho, V. Sangwan, X. Liu, L. Lauhon, T. Marks, and M. Hersam, "Effective Passivation of Exfoliated Black Phosphorus Transistors Against Ambient Degradation," *Nano Lett.*, vol. 14, no. 12, pp. 6964-6970, Nov. 2014. DOI: 10.1021/nl5032293

[13] J. Kim, Y. Liu, W. Zhu, S. Kim, D. Wu, L. Tao, A. Dodabalapur, K. Lai, and D. Akinwande, "Toward Air-Stable Multilayer Phosphorene Thin-Films and Transistors," *Sci. Rep.*, vol. 5, no. 8989, pp. 8989-1-8989-7, Mar. 2015. DOI: 10.1038/srep08989

[14] H. Wang, X. Wang, F. Xia, L. Wang, H. Jiang, Q. Xia, M. Chin, M. Dubey, and S. Han, "Black Phosphorus Radio-Frequency Transistors," *Nano Lett.*, vol. 14, no. 11, pp. 6424-6429, Oct. 2014. DOI: 10.1021/nl5029717

[15] X. Luo, Y. Rahbarihagh, J. Hwang, H. Liu, Y. Du, and P. D. Ye, "Temporal and Thermal Stability of $Al_2O_3$-Passivated Phosphorene MOSFETs," *IEEE Electron Device Lett.,* vol. 35, no. 12, pp. 1314–1316, Dec. 2014. DOI: 10.1109/LED.2014.2362841

[16] M. Buscema, D. Groenedijk, S. Blanter, G. Steele, H. Zant, and A. Gomez, "Fast and Broadband Photoresponse of Few-Layer Black Phosphorus Field-Effect Transistor," *Nano Lett.*, vol. 16, no. 6, pp. 3347-3352, May 2014. DOI: 10.1021/nl5008085

[17] F. Xia, H. Wang, D. Xiao, M. Dubey, and A. Ramasubramaniam, "Two-Dimensional Material Nanophotonics," *Nat. Photo.*, vol. 8, no. 12, pp. 899-907, Nov. 2014. DOI: 10.1038/nphoton.2014.271

[18] Y. Deng, Z. Luo, N. Conrad, H. Liu, Y. Gong, S. Najmaei, P. Ajayan, J. Lou, X. Xu, and P. D. Ye, "Black Phosphorus-Monolayer $MoS_2$ van der Waals Heterojunction P-N Diode," *ACS Nano*, vol. 8, no. 8, pp. 8292-8299, Jul. 2014. DOI: 10.1021/nn5027388

[19] Z. Luo, J. Maassen, Y. Deng, Y. Du, M. Lundstrom, P. D. Ye, and X. Xu, "Anisotropic In-Plane Thermal Conductivity Observed in Few-Layer Black Phosphorus," *Nat. Commun.*, vol. 6, no. 8572, pp. 8572-1-8572-8, Oct. 2015. DOI: 10.1038/ncomms9572

[20] N. Youngblood, C. Chen, S. Koester, and M. Li, "Waveguide-Integrated Black Phosphorus Photodetector with High Responsivity and Low Dark Current," *Nat. Photo.*, Mar. 2015. DOI: 10.1038/nphoton.2015.23

[21] L. Li, F. Yang, G. Ye, Z. Zhang, Z. Zhu, W. Lou, L. Li, K. Watanabe, T. Taniguchi, K. Chang, Y. Wang, X. Chen, and Y. Zhang, "Quantum Hall Effect in Black Phosphorus Two-Dimensional Electron Gas," *arXiv:1504.04731*.

[22] Z. Xiang, G. Ye, C. Shang, B. Lei, N. Wang, K. Yang, D. Liu, F. Meng, X. Luo, L. Zou, Z. Sun, Y. Zhang, and X. Chen, "Pressure-Induced Lifshitz Transition in Black Phosphorus," *arXiv:1504.00125*.

[23] L. Li, G. Ye, V. Tran, R. Fei, G. Chen, H. Wang, J. Wang, K. Watanabe, T. Taniguchi, L. Yang, X. Chen, and Y. Zhang, "Quantum Oscillations in A Two-Dimensional Electron Gas in Black Phosphorus Thin Film," *Nat. Nanotechnol.*, vol. 10, no. 7, pp. 608-613, May 2015. DOI: 10.1038/nnano.2015.91

[24] N. Gillgren, D. Wickramaratne, Y. Shi, T. Espiritu, J. Yang, J. Hu, J. Wei, X. Liu, Z. Mao, K. Watanabe, T. Taniguchi, M. Bockrath, Y. Barlas, R. Lake, and C. Lau, "Gate Tunable Quantum Oscillations in Air-Stable and High Mobility Few-Layer Phosphorene Heterostructures," *2D Mat.*, vol. 2, no. 1, pp. 011001, Dec. 2014. DOI: doi:10.1088/2053-1583/2/1/011001

[25] X. Chen, Y. Wu, Z. Wu, Y. Han, S. Xu, L. Wang, W. Ye, T. Han, Y. He, Y. Cai, and N. Wang, "High Quality Sandwiched Black Phosphorus Heterostructure and Its Quantum Oscillations," *Nat. Commun.*, vol. 6, no. 7315, pp. 7315-1-7315-6, Jun. 2015. DOI: 10.1038/ncomms8315

[26] Y. Cao, A. Mishchenko, G. Yu, K. Khestanova, A. Rooney, E. Prestat, A. Kretinin, P. Blake, M. Shalom, G. Balakrishnan, I. Grigorieva, K. Novoselov, B. Piot, M. Potemski, K. Watanabe, T. Taniguchi, S. Haigh, A. Geim, and R. Gorbachev, "Quality Heterostructure from Two Dimensional Crystals Unstable in Air by Their Assembly in Inert Atmosphere," *Nano Lett.*, vol. 15, no. 8, pp. 4914-4921, Jul. 2015. DOI: 10.1021/acs.nanolett.5b00648

[27] H. Liu, Y. Liu, and D. Zhu, "Chemical Doping of Graphene," *J. Mat. Chem.* vol. 21, no. 10, pp. 3253-3496, Nov. 2011. DOI: 10.1039/C0JM02922J

[28] W. Chen, S. Chen, D. Qi, X. Gao, and A. Wee, "Surface Transfer p-Type Doping of Epitaxial Graphene," *J. Am. Chem. Soc.*, vol. 129, no. 34, pp. 10418-10422, Aug. 2007. DOI: 10.1021/ja071658g

[29] G. Nikiforov, L. Ono, and Y. Qi, "p-Doping of Squaraine with F4-TCNQ by Solution Processing," *ITE Trans. on Med. Tech. Appl.*, vol. 3, no. 2, pp. 133-142, Apr. 2015. DOI: 10.3169/mta.3.133

[30] L. Yu, A. Zubair, E. Santos, X. Zhang, Y. Lin, Y. Zhang, and T. Palacios, "High-Performance $WSe_2$ Complementary Metal Oxide Semiconductor Technology and Integrated Circuits," *Nano Lett.*, vol. 15, no. 8, pp. 4928-4934, Jul. 2015. DOI: 10.1021/acs.nanolett.5b00668